# Fully tunable exciton-polaritons emerging from WS$_2$ monolayer excitons in an optical lattice at room temperature

L. Lackner[1,2,†], M. Dusel[1], O.A. Egorov[3], H. Knopf[4,5,6], F. Eilenberger[4,5,6], S.Schröder[5], S. Tongay[7], C. Anton-Solanas[2], S. Höfling[1] and C. Schneider[1,2,†]

[1]Technische Physik and Wilhelm-Conrad-Röntgen-Research Center for Complex Material Systems, Universität Würzburg, D-97074 Würzburg, Am Hubland, Germany

[2]Institute of Physics, Carl von Ossietzky University, 26129 Oldenburg, Germany.

[3]Institute of Condensed Matter Theory and Solid State Optics, Friedrich Schiller Universität Jena, Max-Wien Platz 1, 07743 Jena, Germany.

[4]Institute of Applied Physics, Abbe Center of Photonics, Friedrich Schiller University, 07745 Jena, Germany.

[5]Fraunhofer-Institute for Applied Optics and Precision Engineering IOF, 07745 Jena, Germany.

[6]Max Planck School of Photonics, 07745 Jena, Germany.

[7]School for Engineering of Matter, Transport, and Energy, Arizona State University, Tempe, Arizona 85287, USA

†Corresponding authors: lukas.lackner@uol.de, christian.schneider@uol.de

**Engineering non-linear hybrid light-matter states in tailored optical lattices is a central research strategy for the simulation of complex Hamiltonians. Excitons in atomically thin crystals are an ideal active medium for such purposes, since they couple strongly with light and bear the potential to harness giant non-linearities and interactions while presenting a simple sample-processing and room temperature operability. We demonstrate lattice polaritons, based on an open, high-quality optical cavity, with an imprinted photonic lattice strongly coupled to excitons in a WS$_2$ monolayer. We experimentally observe the emergence of the canonical band-structure of particles in a one-dimensional lattice at room temperature, and demonstrate frequency reconfigurability over a spectral window exceeding 12 meV, as well as the systematic variation of the nearest neighbour coupling, reflected by a tuneability in the bandwidth of the p-band polaritons by 7 meV. The technology presented in this work is a critical demonstration towards reconfigurable photonic emulators operated with non-linear photonic fluids, offering a simple experimental implementation and working at ambient conditions.**

Exciton-polaritons in optical lattices have matured to a very promising and versatile platform in applied information technology[1–4]. Their bosonic character enables ultra-fast condensation processes[5–8], which are already exploited in classical emulation applications[1,2], and can be also harnessed in ultra-fast quantum annealing architectures[9]. Their intrinsically strong non-linearity has very recently opened the new area of genuine quantum polaritonics[10,11]. The combination of exciton-polaritons with optical lattices, hence, is a powerful approach to the simulation of complex Hamiltonians, with the perspective to explore topologically non-trivial phenomena and possibly quantum effects.

With the emergence of atomically thin crystals and the observation of giant light-matter coupling of excitons hosted in transition metal dichalcogenide (TMDC) monolayers, the research field on exciton-polaritons has experienced a paradigmatic shift on the material side: Excitons in TMDC monolayers present a very large oscillator strength[12] and their large binding energies routinely facilitate the observation of exciton-polaritons from cryogenic up to room temperature[13–16]. At the same time, with the current developments on the material optimization, also their structural homogeneity becomes competitive with high quality epitaxially grown III-V semiconductors[17]. Last but not least, there is a sizeable polariton interaction[18,19] which can be engineered via injecting electrons or combining various monolayers in a van-der-Waals heterostructure [20]. Those advantages, very recently, led to the first observations of polariton condensation in hybrid TMDC-III-V and in pure TMDC microcavity architectures[21,22].

In this work, we significantly expand the versatility of TMDC exciton-polaritons, by demonstrating their formation in a one-dimensional optical lattice imprinted in a so-called open cavity. The structure clearly forms the canonical gapped spectrum of Bloch-polaritons[23–25], and its "on-the-fly" cavity reconfigurability allows us to demonstrate the emergence of frequency tunable Bloch polaritons, with the advantage of operating at room temperature.

**Sample and experimental setup: a room temperature, tunable open cavity**

The implementation of our experimental platform is schematically depicted in Fig. 1(a). Our device is composed of two distributed Bragg reflectors (DBRs) of $SiO_2/TiO_2$ (top and bottom mirror containing 10 and 10 mirror pairs, each), which are separated by an air-gap. Both DBRs are attached to an XYZ set of nano-positioners. The two mirrors form an open cavity, whose cavity resonance is tunable on-demand by varying the DBRs separation with nm-precision. The atomically thin $WS_2$ layer is isolated from a CVD-grown bulk crystal, and transferred onto the bottom DBR via the dry-gel stamping method[26].

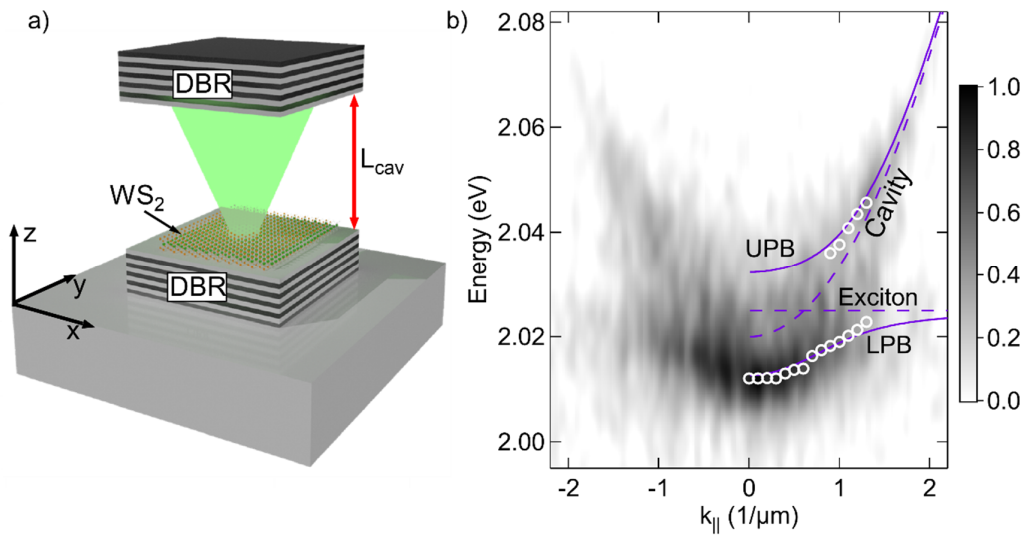

Fig. 1. **Open cavity and TMDC monolayer sketch and strong coupling conditions.** (a) Sketch of the open cavity device, composed of two DBRs embedding an air gap. The $WS_2$ monolayer is placed on top of the bottom DBR. (b) Dispersion relation of the polariton emission for a $L_{cav}$ ~ 2.2 µm, under weak excitation with a green, non-resonant, continuous wave laser. As a result of the peak fitting of the UPB and LPB modes, see white circles, the excitonic and photonic modes [UPB and LPB] are represented with dashed [full] lines for $k_{||}>0$.

As the bottom DBR is terminated on a low index material of $\lambda/(4n)$ thickness ($SiO_2$), the monolayer will permanently remain in the optical field antinode, and thus optimally coupled to the longitudinal optical cavity modes. To facilitate a micrometric approach of the two DBR mirrors (<10 µm), we have etched a rectangular mesa of dimensions 200 µm x 200 µm, and 20 µm depth into the GaAs substrate, before evaporating the bottom DBR[27].

It is possible to assess the physical distance between the two DBR mirrors by analyzing the separation of the longitudinal optical cavity modes and comparing those to transfer matrix simulations (see Suppl. Materials S3). From those studies, we find that our mirrors can be approached to a distance of ~2 µm.

Next, at a mirror distance of 2.2±0.5 µm, we study the light-matter coupling conditions in our device by inspecting the dispersion relation of the polariton photoluminescence under weak excitation with a continuous wave green laser (see Methods for more details), see Fig. 1 (b). The longitudinal resonance mode of the open cavity ($E_C = 2.020$ eV) is spectrally tuned to resonance with the exciton energy of the WS$_2$ monolayer ($E_X = 2.025$ eV), with a negative detuning of $\Delta = E_C - E_X = -5$ meV. At resonance conditions, the emitted cavity photoluminescence deviates from the standard parabolic dispersion relation of an empty cavity and breaks into two upper and lower polariton branches (UPB and LPB). We analyze the peak position of the respective polariton modes by iterative fitting, see white circles in Fig. 1(b). In order to determine the Rabi splitting, we apply a standard coupled oscillator approach (see Suppl. Materials S2), reflecting strong coupling conditions in our device with a splitting of $\Omega = 19 \pm 3$ meV. It is worth noting, that a similar magnitude of the Rabi-splitting has been previously found in TMDC-loaded open cavities of similar effective length[14], and it is in full agreement with our numerical simulations using the transfer matrix method (see Suppl. Materials S3).

**Room temperature TMDC Bloch Polaritons in an open cavity**

To engineer the confinement of TMDC-polaritons, we utilize focused Gallium ion beam (FIB) milling to mechanically shape hemispheric traps of 5 µm diameter and an approximate depth of ~350 nm in a glass carrier; the top DBR is deposited in the last step via dielectric sputtering. It is worth noting, that the sputtering process does not yield a planarization, or a significant roughening of the FIB structure.

Polariton trapping in single hemispheric traps manifests in a discretized dispersion, giving rise to well-defined confined modes labelled s, p, and d in this paper (Fig. S3). In the presence of a neighbouring trap with sufficiently strong overlap, we can witness the formation of molecular modes, reflected by a molecular splitting most notable in the s- and the d-state (Fig. S3). By utilizing this accurate control of mode-coupling, we create one-dimensional lattice structures with varying inter-trap distance (A) and D is the trap diameter(D=5 µm, for the experiments shown here). In the open cavity configuration, this yields tunable optical lattices, as sketched in Fig. 2(a), as well as schematically depicted in the top parts of Fig. 2(b-d).

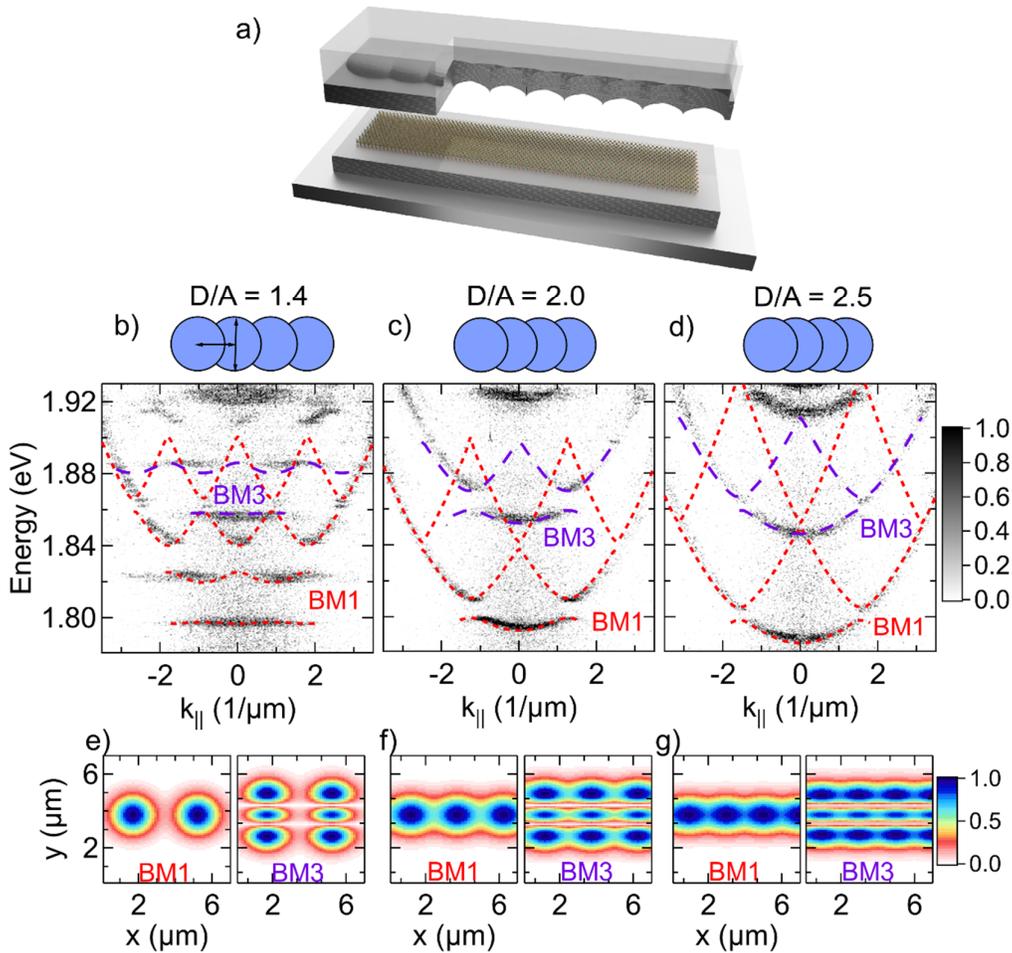

Fig. 2. **Optical properties of TMDC polaritons in a lattice.** a) Sketch of the open cavity device with the 1D photonic potential implemented in the top mirror. (b-d) Dispersion relation of the photoluminescence from one-dimensional polariton lattices (see the sketch of the photonic potential in the top part of each panel) with increasing diameter (vertical arrow) to center-center distance (horizontal arrow) ratio D/A=1.4, 2.0 and 2.5, respectively. The dispersion relation maps are encoded in a normalized, false-colour scale. The dashed lines are theory fittings to the experimental data. (e-g) Profiles of the Bloch modes with one (BM1) or three field maxima (BM3) in transverse to the lattice structure.

Figures 2(b-d) show the polariton dispersion relation via momentum-resolved photoluminescence measurements of three different lattices, with traps of D=5 µm diameter and varying overlap A, D/A=1.4, 2.0, and 2.5, respectively. The full control on the spatial configuration of the open cavity allows studying different photonic potentials with the same monolayer, by laterally displacing the top DBR containing the imprinted lattices at different lateral positions.

The photoluminescence maps clearly reveal the formation of a characteristically gapped massive-particle dispersion relation of polaritons in a lattice. The magnitude of the energy gap, which separates the photonic bands emerging from the coupling of s- and p-type orbitals, as well as the curvature of the bands, critically depends on the nearest-neighbor hopping in the photonic lattice which is determined by the D/A ratio.

While the curvature of the polariton bands strongly increases for increasing nearest-neighbor overlap (D/A), the size of the photonic gap, which emerges at the edge of the one-dimensional Brillouin zone, decreases from 26±2 to 4±2 meV. In order to provide a deeper and more quantitative understanding of our system, we model the dispersion relation as well as the Bloch modes emerging in our system in the framework of a mean-field model with effective potential (dashed lines in Figs. 2(b-d). Details on the model are found in the Supp. Material S5. Our model reveals nearest-neighbor hopping constants of ~0.2 to 2.7 meV, which are of similar magnitude as in previous works utilizing coupled hemispheric cavities with organic materials[28]. The corresponding real-space profile fo Bloch-modes yielding the s-band formation (BM1) is plotted in Fig. 2(e-g)

It is worth noting, that the emergence of a second Bloch band, approx. 50 meV blue-shifted from the ground-mode (at 1.85 meV in Fig. 2b, the theory is plotted in dashed purple lines) is not related to a higher longitudinal mode. It can be associated with the band structure arising from photonic orbitals of d-symmetry, coupling in the π-configuration. The real-space profile of the Bloch-modes (BM3), featuring three maxima in the transverse direction to the one-dimensional lattice, is plotted in Fig. 2(e-g).

In contrast, only the sigma-coupling of the p-orbitals can be seen in our spectra (the second band, dashed red lines).The π-coupling of p-orbitals (Suppl. Fig. S4, BM2), featuring an even number of field maxima in transverse direction to the lattice,  as expected, remain dark for a pump-spot oriented in the center of the lattice and should only significantly contribute to the signal for a strongly misplaced excitation pump. The excellent agreement between experiment and theory finally allows us to directly assess a physical cavity length of ~2.2 µm.

**Tuneablity of the optical polariton lattice**

While Bloch band polaritons in microcavity lattices are widely explored in micro-structured GaAs samples at cryogenic conditions[13], and more recently in organic and perovskite structures at ambient conditions[28,29], the remarkable asset of our approach comes along with the intrinsic space- and spectral-tunability provided by the open cavity advantage. The vertical displacement of the micro-structured top DBR changes the optical resonance conditions on demand and consequently controls the eigen-energies of the Bloch–polaritons.

Figure 3(a) depicts the dispersion relation recorded for two different separations of the top DBR, clearly reflecting the energy shift of all relevant Bloch bands: the negative and positive $k_{||}$ regions in panel (a) correspond to a DBR separation of 2.184 and 2.192 µm, respectively.

The open cavity tuneability allows the continuous spectral modification of the polariton band-structure by more than 12 meV, see Fig. 3(b), where we depict the energy change of the s-energy band versus cavity length $L_{cav}$. In addition, the modification of the cavity length yields significant changes in the coupling configuration. We capture a profound modification of the width of the Bloch-bands, which we study for the sigma-coupled p-band (Bloch mode 1, BM1), as well as the π-coupled d-band (Bloch mode 3, BM3). In Fig. 3(c), we plot the extracted bandwidth of these bands as a function of the cavity length. The bandwidth change reflects a tuneability of the nearest neighbor coupling up to 5 meV by nanometrically displacing the cavity by 16 nm, which is in turn an inspiring result towards the exploration of optomechanically coupled lattices.

The emergence of the tunability of photonic hopping can be naively expected to route in a modification of the mode index, as the cavity length is modified. This yields a much slower trend as captured in our experiment (see Suppl. Material S6). However, in our numerical calculations, we found strong indications for the emergence of resonance yielding a very strong dependence between cavity length and nearest-neighbour hopping on the order of our experimental observations.

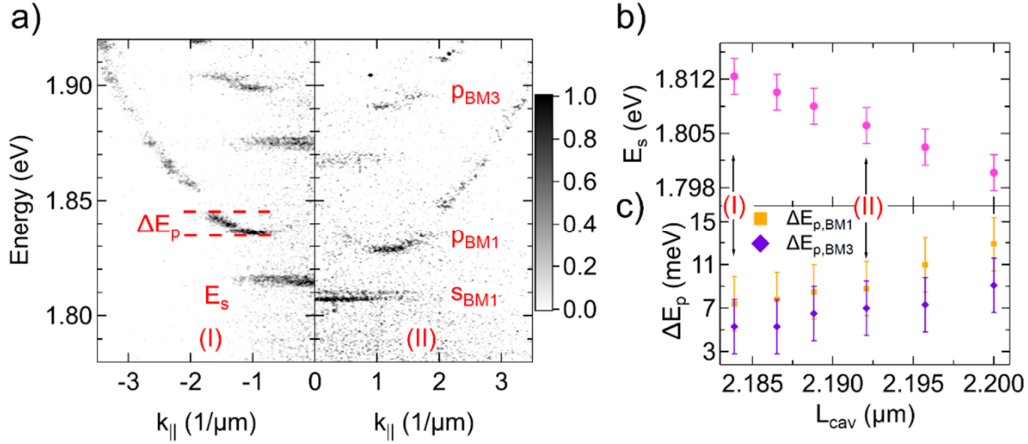

Fig. 3. **Spectral- and coupling-tuning in a polariton lattice.** (a) Dispersion relation maps of a polariton chain with D/A=1.7 for two different DBR separations of 2.184 and 2.192 µm in the regions of $k_\parallel<0$ (I, left) and $k_\parallel>0$ (II, right), respectively. The map intensity is coded in a false colour scale included in the right side of the map. (b,c) Dependence of the s-band energy (magenta circles) and the p-band bandwidth (for Bloch modes 1 (BM1) and 3 (BM3), presenting an energy difference of 7 meV) versus cavity length. The labels (I,II) mark the corresponding $E_s$ and $\Delta E_p$ values of panels in (a).

## Conclusions

We have presented an open cavity system with an integrated optical lattice, strongly coupled to an atomically thin layer of $WS_2$ at room temperature. The imprinted lattice geometries in the open cavity leads to a great control and reconfigurability in the engineering of Bloch-polaritons. We demonstrate a strong spectral tuneability in energy and lattice coupling. While our experiments are carried out in the linear regime, it is reasonable to assume that the non-linear regime of Bosonic condensation can be reached in a similar approach. The possibility to engineer and increase the Kerr non-linearities of TMDC polaritons, e.g. by utilizing van der Waals heterostructures, makes our system highly interesting for applications relying on non-linear photonic lattices.

## Methods

The air-gap open cavity is composed of two distributed Bragg reflectors (DBRs) of $SiO_2/TiO_2$ (top and bottom mirror containing 10 and 10 mirror pairs. In order to increase the minimum approach between the two mirrors, the GaAs substrate of the bottom mirror is etched forming a mesa with dimensions 200 µm x 200 µm, and 20 µm depth. The hemispheric photonic traps of the top mirror are sculpted via FIB in a glass carrier. These traps present a diameter of 5 µm and an approximate depth of ~350 nm. Each DBR is formed via dielectric sputtering. It is worth noting, that the sputtering process in the DBR formation does not yield a planarization, or a significant roughening of the FIB structure in the top mirror.

Each DBR is mounted in two independent XYZ nano-positioners model Attocube ECSxy5050/Al/NUM/RT (for horizontal displacement) and ECSz5050/Al/NUM/RT (for vertical cavity displacement).

The $WS_2$ monolayer (obtained from a CVD-grown bulk crystal) is transferred onto the bottom DBR via the dry-gel stamping method. The thickness of each $SiO_2$ [$TiO_2$] $\lambda/4$-layer is 102.9 [64.6] nm, optimized for a 630 nm cavity resonance. The cavity is terminated in $SiO_2$ (lower refractive index) to ensure the maximum coupling of the $WS_2$ monolayer to the photonic cavity field.

The laser source is a continuous wave, green laser; its pump power is controlled with adjustable optical density filters in the excitation path. The laser is tightly focused in the top mirror of the cavity with an Mitutoyo objective (×50, 0.42 NA). This allows to form a Gaussian spot size with a FWHM diameter of ~3.6 μm. The spectrometer [CCD] used for these experiments is an Andor Shamrock 500i [Andor Newton EMCCD DU971P-BV].

**Acknowledgement**

The authors gratefully acknowledge funding by the State of Bavaria and Lower Saxony. Funding provided by the European Research Council (ERC project 679288, unlimit-2D) is acknowledged. We thank N. Carlon Zambon for his help in the transfer-matrix method simulations, Johannes Michl for the graphics design and A. Chernikov for providing accurate refractive index data for $WS_2$. ST acknowledges the use of facilities within the Eyring Materials Center at Arizona State University supported in part by NNCI-ECCS-1542160. S.T acknowledges support from DOE-SC0020653, NSF CMMI 1933214, NSF DMR 1552220 and DMR 1955889. FE and HK are supported by the Federal Ministry of Education and Science of Germany under Grant ID 13XP5053A.

**Supplementary Materials**

*S1. Longitudinal optical cavity modes versus DBR separation*

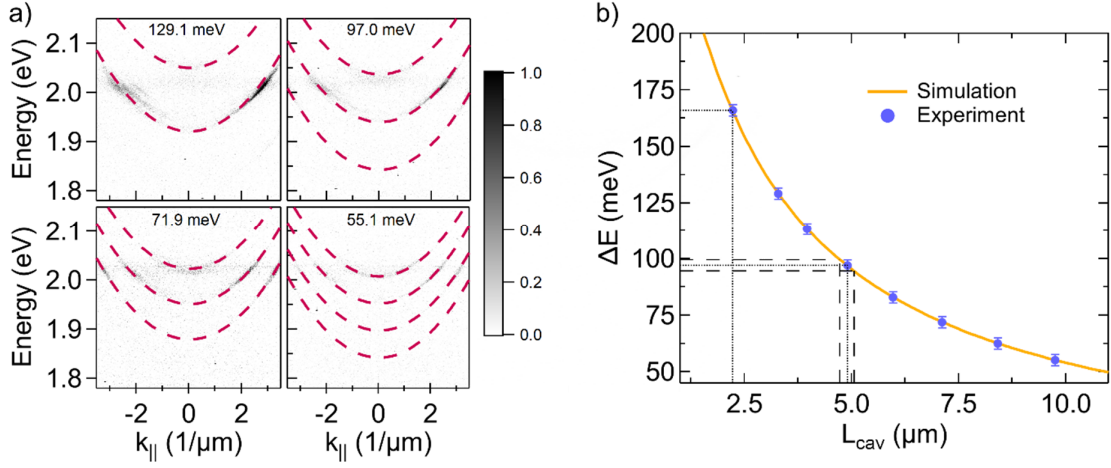

**Fig. S1.** (a) Dispersion relation of the cavity under different DBR separations, 3.3, 4.9, 7.1 and 9.8 µm, respectively. The different parabolic cavity modes are highlighted with dashed lines. The excitation setting is the same as that indicated for Fig. 1(b) of the main text. (b) Transfer Matrix Simulation of the energy splitting between the cavity modes as a function of DBR separation, the violet circles located on the full line depict the different experimental energy splittings (and therefore the different DBR separations) which have been studied.

*S2. Description of the coupled oscillator model*

To describe the upper and lower polariton resonances, we employ a standard two-coupled-oscillators model:

$$\begin{bmatrix} E_{ex} & V/2 \\ V/2 & E_{cav} \end{bmatrix} \begin{bmatrix} X \\ C \end{bmatrix} = E \begin{bmatrix} X \\ C \end{bmatrix} \quad (S.1.1)$$

where $E_{ex}$ and $E_{cav}$ denote the energies of the exciton and cavity modes, respectively, and $V$ is the normal mode splitting. For the lower polariton branch the Hopfield coefficients $X$ and $C$ are given by:

$$|X|^2 = \frac{1}{2}\left(1 + \frac{E_{cav} - E_{exc}}{\sqrt{(E_{cav} - E_{exc})^2 + V^2}}\right) \quad (S.1.2)$$

$$|C|^2 = 1 - |X|^2 \quad (S.1.3)$$

Their squared amplitudes $|X|^2$ and $|C|^2$ quantify the exciton and cavity photon fractions. The eigenenergies of the upper and lower polariton branches are obtained by solving the eigenvalue problem:

$$E_{UP,LP}(k_{||}) = \frac{1}{2}\left(E_{ex} + E_{cav} \pm \sqrt{V^2 + (E_{cav} - E_{exc})^2}\right) \quad (S.1.4)$$

## S3. Transfer Matrix Method for the simulation of the dispersion relation

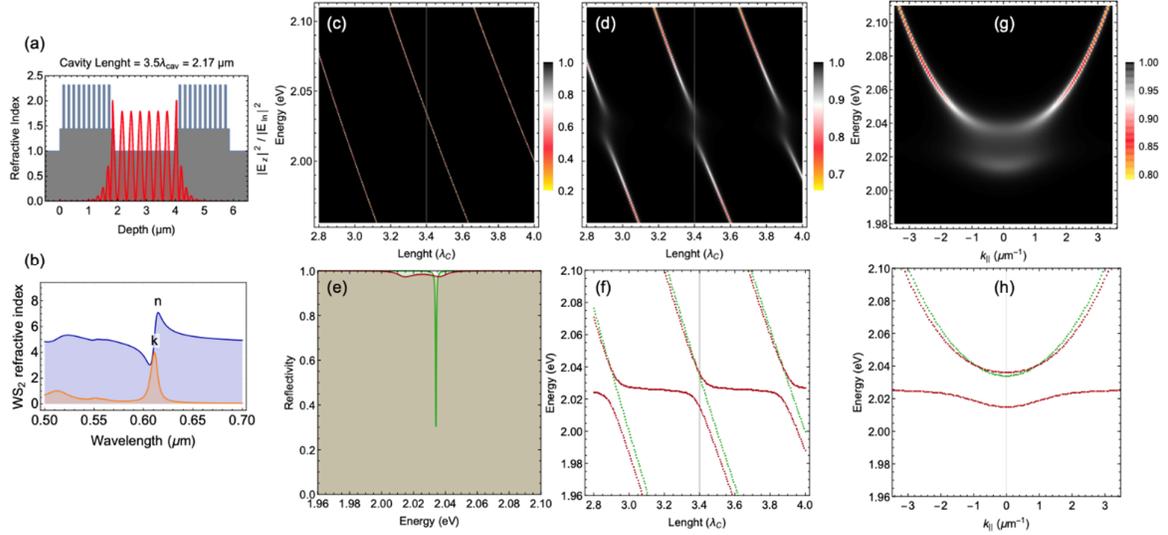

**Fig. S2.** (a) Refractive index profile of the cavity (left axis, gray color) and normalised intensity profile of the cavity mode at the cavity resonance $\lambda_C = 0.62 \ \mu m$ (right axis, red color) versus cavity depth. The air spacer has been simulated for a cavity length of $L_C = \lambda_C$. The cavity configuration presents a maximum in the place where the WS$_2$ monolayer is deposited (bottom mirror surface). (b) WS$_2$ monolayer refractive index versus wavelength, extracted from [30]. (c/d) Reflectivity map versus energy (vertical axis) and cavity length (horizontal, in units of $\lambda_C = 0.62 \ \mu m$) in absence/presence of a monolayer of WS$_2$ (deposited on the bottom mirror, with a thickness of 0.65 nm). The grey vertical line at $L_C = 3.4\lambda_C$ in these panels represents the corresponding reflectivity spectrum represented in panel (e), where the green/red colour trace encodes the reflectivity spectrum in absence/presence of the WS$_2$ monolayer. (f) Extracted from panels (c,d), reflectivity minima versus energy and cavity length, revealing the polariton mode splitting (red trace) occurring from the strong coupling between the longitudinal cavity modes and the WS$_2$ exciton. The green trace corresponds to panel (a), where the WS$_2$ monolayer is absent. The cuts of the grey vertical line with the red trace at $L_C = 3.4\lambda_C$ reveals a Rabi splitting of 20 meV, in excellent agreement with the experimental observations. (g) Reflectivity map versus energy (vertical axis) and in-plane momentum (horizontal), revealing the polariton dispersion, this simulation is obtained for $L_C = 3.4\lambda_C$, similar as that reported in the experimental dispersion relation shown in Fig. 1 of the main text. (h) Extracted from panel (g) and from the reflectivity of the empty cavity (not shown here), reflectivity minima versus energy and cavity length, revealing the dispersion relation of the upper and lower polariton modes (red trace) and the parabolic dispersion relation of the empty cavity (green trace).

## S4. Experimental results on single- and double-trap structures

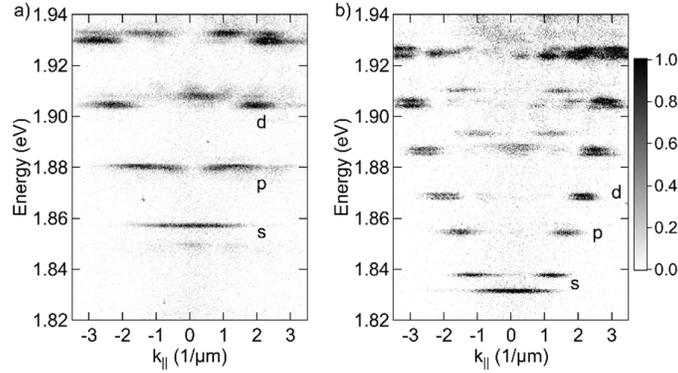

**Fig. S3.** (left/right) Polariton photoluminescence map versus energy and in-plane momentum, showing the dispersion relation of a single lens (left, 6 μm diameter) and a molecule (right, 5.5μm diameter and overlap D/A=2.5). The photoluminescence intensity is encoded in a false colour scale (see normalised scale on the right side). The excitation conditions are the same as those described in the main text.

## S5. Bloch Modes within the mean-field model with effective potential

In order to determine the energy-momentum band structure of polaritons in the 1D lattice, we calculate first the Bloch modes in the mean-field approximation, where the geometry of the chain is represented by an effective potential in two transverse (x and y) directions. The mean-field approach is valid in the vicinity of a longitudinal resonance and it requires that the respective longitudinal mode profile between the mirrors (z-direction) is fixed. Then, in the first approximation of the perturbation theory, it is possible to reduce the three-dimensional problem to respective two-dimensional one (x- and y-) by separating the longitudinal mode profile (z-direction).

Appling this mean-field approach we solve the following eigenvalue problem for the energy $E(k_{||})$ of the Bloch mode with the Bloch vector $\boldsymbol{k} = k_{||}\vec{e}_x$:

$$E(k_{||}) \begin{Bmatrix} p_b(\mathbf{r}, k_{||}) \\ e_b(\mathbf{r}, k_{||}) \end{Bmatrix} = \hat{L}(k_{||}) \begin{Bmatrix} p_b(\mathbf{r}, k_{||}) \\ e_b(\mathbf{r}, k_{||}) \end{Bmatrix}, \tag{S5.1}$$

where the functions $p_b(\mathbf{r}, k_{||})$ and $e_b(\mathbf{r}, k_{||})$ describe the amplitude distributions of the photonic and excitonic components of the Bloch modes in real space defined by the resonator $\mathbf{r} = \{x, y\}$. The main matrix in Eq. (S5.1), describing the single-particle coupled states of excitons and photons, is given by the expression

$$\hat{L}(k_{||}) = \begin{pmatrix} \omega_C^0 + V(\mathbf{r}) - \frac{\hbar}{2m_C}(\vec{\nabla}_\perp + ik_{||}\vec{e}_x)^2 & \Omega_R \\ \Omega_R & \omega_E^0 - \frac{\hbar}{2m_E}(\vec{\nabla}_\perp + ik_{||}\vec{e}_x)^2 \end{pmatrix}. \tag{S5.2}$$

In the model above, the quantities $\omega_C^0$ and $\omega_E^0$ represent the energies of bare photons and excitons ($\hbar\omega_E^0 = 2.025\ eV$), respectively. The photon-exciton coupling strength is given by the parameter $\Omega_R$ which defines the Rabi splitting $2\hbar\Omega_R = 20 meV$ between coupled excitons within TMDC and photons of the cavity mode. The kinetic energy of polaritons is characterized by the effective mass $m_C \approx 5.1 \cdot 10^{-6} m_e$ ($m_e$ free electron mass) which defines transport properties of the intracavity photons. The effective exciton mass is $m_E = 10^5 m_C$.

The periodic array with the period $a$ is modeled by the two-dimensional potential $V(\mathbf{r})$ and consists of spatially-overlapping polaritonic traps. A single, separate trap has an ellipsoid-shaped potential profile defined as the real part of the function:

$$V(x,y) = V_0 \left( \sqrt{1 - \left(\frac{x}{R_x}\right)^2 - \left(\frac{y}{R_y}\right)^2} - 1 \right) \Big/ \left( \sqrt{1 - \left(\frac{d}{2R_y}\right)^2} - 1 \right), \qquad (S5.3)$$

with the potential depth $V_0 = 170\ meV$ and the spatial size $d$. $R_x$ and $R_y$ describe the radii of the hemispheric dimples in the upper mirror which can be slightly different in both transverse directions.

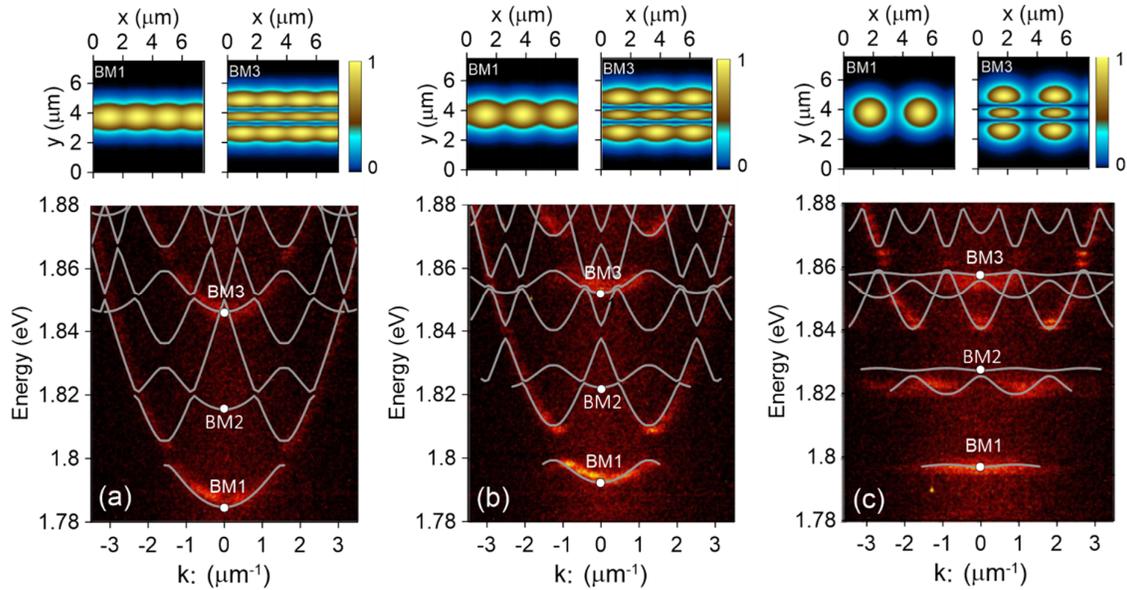

**Fig. S4.** (a-c) Comparison of the theoretical results (solid lines) and experimental measurements (colour map) of the energy-momentum spectrum for three different chain periods, namely, for $2\mu m$ (a), $2.5\mu m$ (b) and $3.5\mu m$ (c). The insets show profiles of the Bloch modes with one (BM1) or three field maxima (BM3) in transverse to chain (y) direction. The best theoretical fits have been obtained for the following system parameters: (a) $\hbar^2/2m_C = 7.2\ meV\cdot\mu m^2$, $\hbar(\omega_C^0 - \omega_E^0) = 261\ meV$, $d = 4.5\mu m$, $R_x = 19.8\mu m$, $R_y = 14.4\mu m$; (b) $\hbar^2/2m_C = 7.5\ meV\cdot\mu m^2$, $\hbar(\omega_C^0 - \omega_E^0) = 255.3\ meV$, $d = 4.8\mu m$, $R_x = 20.2\mu m$, $R_y = 15.4\mu m$; (c) $\hbar^2/2m_C = 8.0\ meV\cdot\mu m^2$, $\hbar(\omega_C^0 - \omega_E^0) = 255.3\ meV$, $d = 4.8\mu m$, $R_x = 18.2\mu m$, $R_y = 15.4\mu m$.

Fig. S4 shows energy-momentum spectra of the Bloch-modes calculated for three different periods of the chain (see also main text, Fig. 2). The higher order modes in transverse to the chain (y-) directions form its own sub-bands, depicted by BM1 and BM3. The respective mode profiles are shown in insets to Fig. S4. It is worth mentioning that, due to the spatial symmetry, the modes with odd number of field maxima only can be visible in the present experiment.

## S6. Bloch Modes beyond mean-field model

The mean-field approximation with an effective potential discussed above allows for calculation of mode profiles in the transverse plane of the cavity (x- and y- directions). However this approach becomes inaccurate if several longitudinal modes (z-direction) are involved into dynamics. For instance this model describes inadequately the cavity-length dependency of the p-band width, discussed in the main text of the manuscript.

To overcome this discrepancy we prepare further mode analysis based on a direct calculation of Maxwell equations in the given refractive index environment, including substrate, Bragg reflectors and the cavity. To avoid unnecessary computational efforts we consider a two-dimensional geometry assuming that the system is homogeneous in the transverse to the chain (y-) direction.

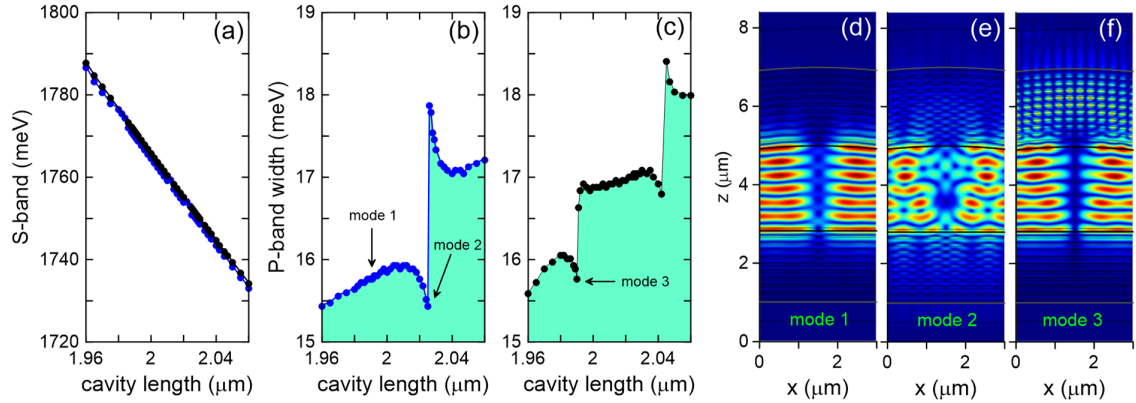

**Fig. S5.** Bloch modes of the open cavity calculated beyond the mean-field approximation (FEM-method). (a) The energy of the bottom of the S-band versus the separation between upper and lower Bragg mirrors (cavity length $L$.) for the chain period $a = 3\mu m$. Panels (b) and (c) show the widths of the p–band depending on the cavity length for TM- (b) and TE- (c) polarised modes, respectively. The width of P-band is defined as the energy difference between modes in the middle ($k_\parallel = 0$) and at the edge ($k_\parallel = \pi/a$) of the Brillouin zone. Panels (d), (e) and (f) show the Bloch mode profiles at the edge of the Brillouin zone for different cavity lengths. (d) TM-polarized mode (y-component of the electric field is zero) for $L = 1.99 \mu m$. (e) TM-polarized mode for $L = 2.025 \mu m$. (f) TE-polarised mode (electric field polarized in y-direction) for $L = 1.99 \mu m$.

By using the finite-element method (FEM, COMSOL Multiphysics) we calculated the energy of s-band versus the cavity length, shown in Fig. S5 (a). As expected the energy of the s-band experiences a red shift of about $43\ meV$ by increasing the cavity length by $100\ nm$. Then we define the width of the p-band as difference between energies of the p-band Bloch modes in the middle and at the edge of the Brillouin zone. The calculated dependence of the P-band width on the cavity length is plotted in Figs. S5, for TM- (b) and TE- (c) polarized modes.

The electric-field profile of the typical p-band Bloch mode (TM) is plotted in Fig. S5(d). Further numerical analysis shows that the respective cavity mode can couple to the Bragg mirror modes resulting in their hybridization (see mode profiles in Figs. S5(e) and (f)). Such interaction between cavity and mirror Bloch modes explains the strong jumps in the p-band width within a small range of modification of the cavity length (see the peaks in Figs. S5(b) and (c)). We believe that this mode hybridization mechanism can explain the observed drop of the p-band width over more than 5 meV by changing the cavity length in the range of several nanometers (see Fig. 3 in the main text of the manuscript).